\documentclass[prl,twocolumn,preprintnumbers,amsmath,amssymb,showpacs]{revtex4}

\usepackage{color}
\usepackage{graphicx}

\def\be{\begin{equation}}
\def\ee{\end{equation}}
\def\l{\lambda}
\def\Tr{{\rm Tr}}

\def\a{\alpha}

\def\bea{\begin{eqnarray}}
\def\eea{\end{eqnarray}}
\begin{document}

\title{Entanglement negativity in quantum field theory} 
\author{Pasquale Calabrese${}^{1}$, John Cardy${}^{2}$, and  Erik Tonni${}^{3}$}

\affiliation{
$^{1}$Dipartimento di Fisica dell'Universit\`a di Pisa and INFN,
56127 Pisa, Italy,\\ 
$^2$The Rudolf Peierls Centre for Theoretical Physics, Oxford
University, Oxford OX1 3NP, UK, and All Souls College, Oxford,\\
$^3$SISSA and INFN,  via Bonomea 265, 34136 Trieste, Italy.}

\date{\today}

\begin{abstract}

 We develop a systematic method to extract the negativity in the ground state of a 1+1 dimensional relativistic quantum 
 field theory, using a path integral formalism to construct the partial transpose $\rho_A^{T_2}$ 
 of the reduced density matrix of a subsystem $A=A_1\cup A_2$, 
 and introducing a replica approach to obtain its trace norm which gives the logarithmic negativity ${\cal E}=\ln ||\rho_A^{T_2}||$. 
 This is shown to reproduce standard results for a pure state. 
 We then apply this method to conformal field theories, deriving the result ${\cal E}\sim(c/4)\ln\big(\ell_1\ell_2/(\ell_1+\ell_2)\big)$ 
 for the case of two adjacent intervals of lengths $\ell_1,\ell_2$ in an infinite system, where $c$ is the central charge. 
 For two disjoint intervals it depends only on the harmonic ratio of the four end points and so is manifestly scale invariant. 
 We check our findings against exact numerical results in the harmonic chain.

\end{abstract}


\maketitle

Recent years have witnessed a large effort to understand and quantify the entanglement content 
of many-body quantum systems (see \cite{rev} for reviews).
This is usually achieved by partitioning an extended quantum system into two 
complementary subsystems and calculating the entanglement entropy $S_A$, 
defined as the von Neumann entropy of the reduced density matrix $\rho_A$ of one subsystem. 
However, this procedure does not give information about the entanglement between 
two non-complementary parts $A_1$ and $A_2$ of a larger system because generically their union is in a mixed state. 
The mutual information $S_{A_1}+S_{A_2}-S_{A_1\cup A_2}$
measures the correlations between the two parts, but gives only an upper bound on the entanglement between them.

A more useful measure of entanglement in this case is the negativity \cite{vw-01}, 
defined as follows. Denoting by $|e_i^{(1)}\rangle$ and $|e_j^{(2)}\rangle$ two bases 
in the Hilbert spaces  ${\cal H}_1$ and ${\cal H}_2$ of each part, 
one first defines the partial transpose 
of  $\rho$ as 
$\langle e_i^{(1)} e_j^{(2)}|\rho^{T_2}|e_k^{(1)} e_l^{(2)}\rangle=\langle e_i^{(1)} e_l^{(2)}|\rho| e^{(1)}_k e^{(2)}_j\rangle$
and then the logarithmic negativity as
\be
{\cal E}\equiv\ln ||\rho^{T_2}||=\ln \Tr |\rho^{T_2}|\,,
\ee
where the trace norm  $||\rho^{T_2}||$ is
the sum of the absolute values of the eigenvalues $\lambda_i$ of $\rho^{T_2}$.
When the two parts are  two microscopic degrees of freedom (e.g. spins),
the negativity coincides with other commonly used entanglement estimators \cite{rev,fazio},
but its definition is more appealing because it is basis independent and so calculable 
by quantum field theory (QFT).

The use of QFT naturally unveils  universal features, in particular close to a quantum critical point. 
For 1D critical theories, that at low energy are also Lorentz invariant, 
the powerful tools of conformal field theory (CFT) can be applied. 
As a matter of fact, 
the interest in
entanglement in extended systems has been considerably boosted by 
the now classical CFT result  that the entanglement entropy
of a large block of length $\ell$ is  $S_A=\frac{c}3 \ln \ell$, with $c$
the {\it central charge}   \cite{Holzhey,Vidal,cc-04}. 
When a subsystem consists of two blocks, the entanglement entropy can also be obtained 
from CFT \cite{fps-08,cct-09,cct-11}, but this gives only the mutual information, not the 
entanglement between the two blocks.

For these reasons, and also motivated by recent results in 
some 1D models \cite{Neg1,Neg2,Neg3}, in this Letter we carry out a systematic 
study of the logarithmic negativity in QFT (in particular  CFT) based
on a new replica formalism.

\begin{figure}[b]
\includegraphics[width=.48\textwidth]{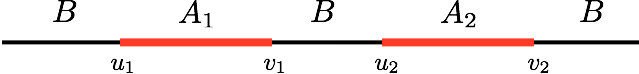}
\caption{We consider the entanglement between two blocks $A_1$ and $A_2$ embedded 
in the ground-state of a larger system. 
}
\label{intervals}
\end{figure}

{\it A replica approach}.
We consider the traces of integer powers of $\rho^{T_2}$.  
For $n$ even (odd), let us say $n_e=2m$ ($n_o=2m+1$), they read
\bea
\Tr (\rho^{T_2})^{n_e}&=&\sum_i \lambda_i^{n_e}= \sum_{\l_i>0} |\l_i|^{n_e}+ \sum_{\l_i<0} |\l_i|^{n_e}\,,\\
\Tr (\rho^{T_2})^{n_o}&=&\sum_i \lambda_i^{n_o}= \sum_{\l_i>0} |\l_i|^{n_o}- \sum_{\l_i<0} |\l_i|^{n_o}\,.\nonumber
\eea
The analytic continuations from even and odd $n$ are different and 
 the trace norm in which we are interested is obtained
by considering the analytic continuation of the even sequence at $n_e\to1$, 
i.e. $\displaystyle {\cal E}=\lim_{n_e\to1} \ln \Tr (\rho^{T_2})^{n_e}$, while the limit $n_o\to1$ 
gives the normalization $\Tr \rho^{T_2}=1$.

As a first example, let us consider  the case in which $\rho=|\Psi\rangle\langle\Psi|$
corresponds to a pure state $|\Psi\rangle$.
Then, the eigenvalues of $\rho^{T_2}$ are related to 
the Schmidt decomposition coefficients \cite{vw-01,lev} 
and after simple algebra
\be
\Tr (\rho^{T_2})^{n_e}= (\Tr \rho_{2}^{n_e/2})^2\,,\label{pure}\qquad 
\Tr (\rho^{T_2})^{n_o}=\Tr \rho_{2}^{n_o}\,,
\ee
where $\rho_2$ is the reduced density matrix on ${\cal H}_2$. 
Taking the limit  $n_e\to 1$, we recover the result \cite{vw-01}
that for a pure state
the logarithmic negativity is the R\'enyi entropy $S_{1/2}=2\ln\Tr\rho_2^{1/2}$.

\begin{figure}[t]
\includegraphics[width=.48\textwidth]{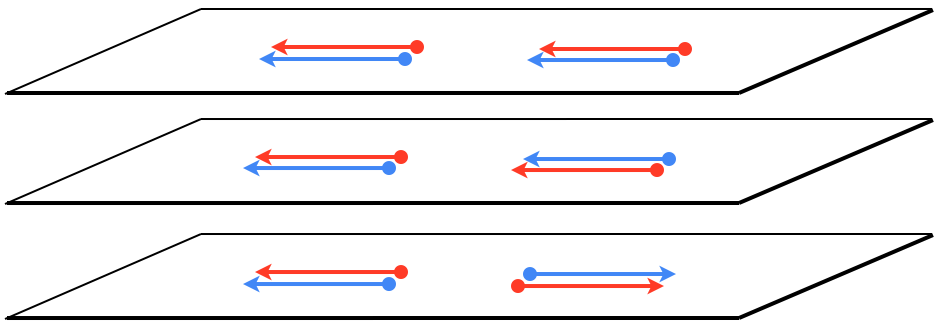}
\caption{Top: The reduced density matrix $\rho_A$ of two disjoint intervals. 
Middle: Partial transpose with respect to the second interval $\rho_A^{T_{2}}$.
Bottom:  Reversed partial transpose $\rho_A^{C_{2}}$.
}
\label{rhos}
\end{figure}

\begin{figure}[t]
\includegraphics[width=.48\textwidth]{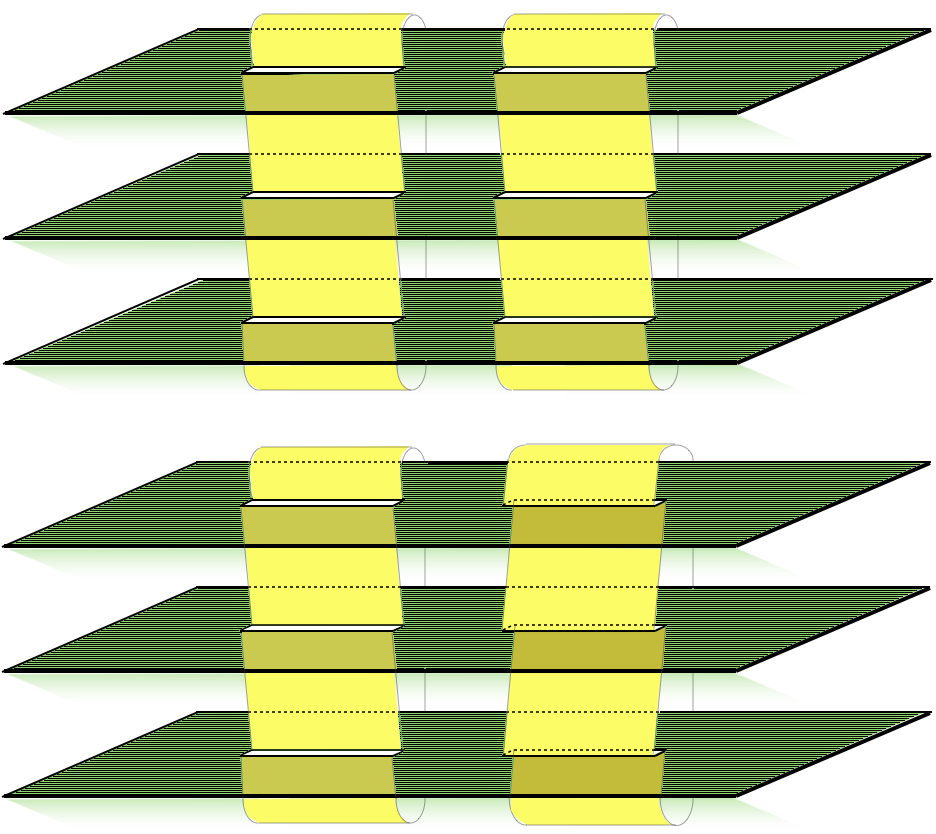}
\caption{Path integral representation of $\Tr\rho_A^n$ (top) and 
$\Tr(\rho_A^{T_{2}})^n$ (bottom) for $n=3$.
}
\label{replicas}
\end{figure}

{\it Negativity and QFT}.
For concreteness we  refer to a 1D system and we consider the {\it tripartition} 
depicted in Fig.~\ref{intervals} with $A$ composed of two parts $A=A_1\cup A_2=[u_1,v_1]\cup [u_2,v_2]$
and $B$ the remainder, but most of the following ideas apply to more general cases.  
In the ground-state of a  QFT, the reduced density matrix $\rho_A$ has the path integral representation 
in Fig.~\ref{rhos} (top)  \cite{cc-04}.  
The two open cuts correspond to the rows and columns of $\rho_A$.
$\Tr \rho_A^n$ for integer $n$ can be obtained by joining cyclically $n$ of the above density matrices as in 
Fig.~\ref{replicas} (top). 
Thus $\Tr\rho_A^n$ is (proportional to) the partition function on this $n$-sheeted Riemann surface
which is equivalent to the correlation 
function of the {\it twist fields} ${\cal T}_n(z)$ constructed exploiting the cyclic permutation symmetry of the sheets, i.e.
\cite{cc-04,cct-09}
\be
\Tr\rho_A^n=\langle {\cal T}_n(u_1)\bar{\cal T}_n(v_1) {\cal T}_n(u_2)\bar{\cal T}_n(v_2)\rangle\,.
\label{rhoatw}
\ee

The partial transposition with respect to  the second interval $A_2$ 
corresponds to the exchange of  row and column indices in $A_2$.  
In the path integral representation, this is equivalent to interchange the upper and lower edges of the second cut
in $\rho_A$ as in the middle of Fig.~\ref{rhos}.
It is convenient to reverse the order of the column and row indices in $A_2$
as in the bottom of Fig.~\ref{rhos}, to obtain the {\it reversed partial transpose} $\rho_A^{C_{2}}$.
This is related to the partial transpose as $\rho_A^{C_{2}}=C \rho_A^{T_{2}} C$, 
where $C$ reverses the order of indices either on the lower or on the upper cut. 
Clearly 
$\Tr(\rho_A^{T_{2}})^n=\Tr(\rho_A^{C_{2}})^n$ 
and so $\Tr(\rho_A^{T_{2}})^n$ is the partition function on the $n$-sheeted surface obtained by joining 
cyclically $n$ of the above $\rho_A^{C_{2}}$ as in the bottom of Fig.~\ref{replicas}.
It is then straightforward to see that 
\be
\Tr(\rho_A^{T_{2}})^n=\langle {\cal T}_n(u_1)\bar{\cal T}_n(v_1) \bar{\cal T}_n(u_2){\cal T}_n(v_2)\rangle\,,
\label{4ptdef}
\ee
i.e. the partial transposition has the net effect to exchange two twist operators compared to Eq. (\ref{rhoatw}).
To replace $\rho_A^{T_{2}}$ with $\rho_A^{C_{2}}$ it has been fundamental to 
consider integer cyclical traces. The operator $C$ enters in quantities like  
$\Tr(\rho_A\rho_A^{T_{2}})$ 
which is in fact the partition function on a non-orientable surface with the topology of a Klein bottle.
This can be computed using CFT methods \cite{klein}.

For $n=2$, ${\cal T}_2=\bar{\cal T}_2$ and so $\Tr\rho_A^2=\Tr(\rho_A^{T_{2}})^2$ 
which follows from the properties of the trace.

We first specialize to a pure state by letting $B\to \emptyset$ for which 
$\Tr (\rho_A^{T_2})^{n}$ can be worked out in full generality as
\be
\Tr (\rho_A^{T_2})^{n}=\langle {\cal T}^2_{n}(u_2) \bar{\cal T}^2_{n}(v_2)\rangle\,.
\ee
This expression depends on the parity of $n$ because  ${\cal T}_n^2$ connects the $j$-th sheet 
with the $(j+2)$-th one. 
For $n=n_e$ even, 
the $n_e$-sheeted Riemann surface decouples in two independent ($n_e/2$)-sheeted 
surfaces. 
Conversely for $n=n_o$ odd, 
the surface remains a 
$n_o$-sheeted Riemann
surface. 
Thus we have
\bea
\Tr (\rho_A^{T_2})^{n_e}&=& (\langle {\cal T}_{n_e/2}(u_2) \bar{\cal T}_{n_e/2}(v_2)\rangle)^2=(\Tr\rho_{A_2}^{n_e/2})^2\,,\nonumber \\
\Tr (\rho_A^{T_2})^{n_o}&= & \langle {\cal T}_{n_o}(u_2) \bar{\cal T}_{n_o}(v_2)\rangle=\Tr \rho_{A_2}^{n_o}\,,
\eea
which are the results for pure states in Eq. (\ref{pure}), recovered here purely from QFT.

We now specialize to the case of a CFT, for which  the twist fields transform like primary operators of dimension
$\Delta_{{\cal T}_n}=c (n-1/n)/12$ \cite{cc-04}.
Thus when $A_2$ is embedded in an infinite system we have ($\ell=u_2-v_2$)
\be
\Tr (\rho_A^{T_2})^{n_e}
\propto
{\ell ^{-\frac{c}{3}(\frac{n_e}2-\frac2{n_e})}},\;\;
\Tr (\rho_A^{T_2})^{n_o}\propto 
{\ell ^{-\frac{c}6(n_o-\frac1{n_o})}}.
\label{1int}
\ee
Despite of the simplicity of the above calculation, it shows one important point of the CFT analysis:
for $n=n_e$ even, 
${\cal T}^2_{n_e}$ has dimension $\Delta_{{\cal T}_{n_e}^2}=c (n_e/2-2/n_e)/6$, while 
for $n=n_o$ odd,  ${\cal T}^2_{n_o}$ has dimension $\Delta_{{\cal T}_{n_o}^2}=c(n_o-1/n_o)/12$,
the same as ${\cal T}_{n_o}$.  
We finally have 
\be
|| \rho_A^{T_2}||= \lim_{n_e\to1} \Tr (\rho_A^{T_2})^{n_e}\propto
 \ell^{\frac{c}2} \Rightarrow\; {\cal E}=\frac{c}2\ln \ell+{\rm cnst}\,.
\label{neg2pt}
\ee

{\it Two adjacent intervals}.
Let us now consider  the non-trivial configuration in which two intervals $A_1$ and $A_2$ of length $\ell_1$ and $\ell_2$ 
share a common boundary (let us say at the origin) which is described by the  3-point function
\be
\Tr (\rho_A^{T_{2}})^{n}=\langle {\cal T}_{n}(-\ell_1) \bar{\cal T}^2_{n}(0){\cal T}_{n}(\ell_2) \rangle\,,
\ee
whose form is determined by conformal symmetry \cite{cft-book}.
For $n=n_e$ even, using the dimensions of the twist operators calculated above, we find
\be
\Tr (\rho_A^{T_{2}})^{n_e}\propto 
{(\ell_1\ell_2)^{-\frac{c}6(\frac{n_e}2-\frac2{n_e})} (\ell_1+\ell_2)^{-\frac{c}6(\frac{n_e}2+\frac1{n_e})}}\,,
\label{3pteven}
\ee
that in the limit $n_e\to1$ gives
\be
|| \rho_A^{T_{2}}||\propto \left(\frac{\ell_1 \ell_2}{\ell_1+\ell_2}\right)^{\frac{c}4}\Rightarrow\;
{\cal E}= \frac{c}4 \ln \frac{\ell_1\ell_2}{\ell_1+\ell_2}+ {\rm cnst}.
\ee
For $n=n_o$ odd, 
$\Tr (\rho_A^{T_{2}})^{n_o} \propto (\ell_1\ell_2(\ell_1+\ell_2))^{-\frac{c}{12}(n_o-\frac{1}{n_o})} $
that for $n_o\to 1$ gives again $\Tr \rho_A^{T_{2}}=1$. 

All the previous results may be generalized to the case of a finite system by using a conformal 
mapping from the cylinder to the plane.
This results in replacing $\ell$ with the chord length $(L/\pi) \sin(\pi \ell/L)$.

{\it Two disjoint intervals}.
For the more interesting and complicated situation of two disjoint intervals of
Fig.~\ref{intervals}, global conformal invariance gives ($\ell_i=|v_i-u_i|$)
\be
\Tr (\rho_A^{T_{2}})^{n}\propto
{[{\ell_1\ell_2(1-y)}]^{-\frac{c}6(n-\frac{1}n)}}  {\cal G}_{n}(y) \,,
\label{Gn}
\ee
where $y=\frac{(v_1-u_1)(v_2-u_2)}{(u_2-u_1)(v_2-v_1)}$ 
is the four-point ratio ($0<y<1$) and ${\cal G}_n(y)$ a function depending on the full operator 
content of the theory. 
$\Tr\rho_A^n$ in Eq. (\ref{rhoatw}) admits the same scaling form, but with a 
different scaling function ${\cal F}_n(y)$  which has 
been calculated for the free compactified boson and for the Ising model \cite{fps-08,cct-09,cct-11}. 
Since Eqs. (\ref{rhoatw}) and (\ref{4ptdef}) are related by an exchange of two twist fields, 
these two functions are related as 
\be
{\cal G}_{n}(y)= (1-y)^{\frac{c}3\left(n-\frac{1}n\right)}{\cal F}_{n}\big(y/({y-1})\big)\,.
\label{GvsF}
\ee
Taking the replica limit $n_e\to1$, we obtain 
\be
{\cal E}(y)=\lim_{n_e\to1} \ln {\cal G}_{n_e}(y)=
\lim_{n_e\to1} \ln \big[{\cal F}_{n_e} \big({y}/({y-1})\big)\big]\,.
\label{NEG}
\ee
Then for conformal invariant systems, the negativity is a scale invariant quantity 
(i.e. a function only of $y$)
because all the dimensional prefactors cancel in the replica limit. 
This has been argued already in the literature on the basis of numerical data \cite{Neg1,Neg2}, but never proved. 

In Refs. \cite{cct-09,cct-11} the function ${\cal F}_{n}(x)$ has been obtained for some CFTs 
only for $0<x<1$ and it is a non-trivial 
technical problem to extend it to the domain $x<0$ in which we are now interested. 
It is a hard open problem to find the analytic continuation to $n_e\to1$.
We will report these technicalities for few specific cases elsewhere \cite{cct-prep} and we limit here to 
discussing the main physical consequences of Eqs.~(\ref{Gn}), (\ref{GvsF}), and~(\ref{NEG}).
These are highlighted by considering the limit $y\to1$ and $y\to0$, i.e.  close and  far intervals respectively.
If $u_2\to v_1$ then $y\to1$ and we should recover the previous result for adjacent intervals. 
Comparing Eqs. (\ref{3pteven}) and (\ref{Gn}) we have ${\cal G}_n(y)\propto (1-y)^\a$ (apart from possible 
multiplicative logarithmic corrections)
with $\a$ equal to 
$\Delta_{{\cal T}_{n}^2}$ the dimension of ${\cal T}_{n}^2$, 
i.e. $\a_{n_e}=c(n_e/2-2/n_e)/6$ and $\a_{n_o}=c(n_o-1/n_o)/12$.
For $n_e\to1$ we have $\a_{n_e\to1}=-c/4$, i.e. the scaling function {\it diverges} approaching $y=1$. 
The opposite limit of  far intervals $y\to 0$ is worked out from the small $y$ expansion of 
${\cal F}_n(y)$ carried out in full generality in Ref.~\cite{cct-11}.
This is a sum over all intermediate operators of the form ${\cal F}_n(y)=\sum_{i} y^{2\Delta_i} s_n(i)$.
The  coefficients $s_n(i)$ have been explicitly calculated \cite{cct-11} and they do not depend on the parity of $n$. 
Thus, in the limit $n\to1$ all these coefficients vanish, because the analytic 
continuation for even and odd $n$ is the same (as the direct computation shows)
and ${\cal E}(y)$ vanishes in $y=0$ faster than any power. 
 %

{\it The harmonic chain}.
We check the CFT results against exact computations in the harmonic chain with Hamiltonian  
\begin{equation}
\label{hamiltonian chain}
H=\frac12 \sum_{j=1}^L\left[
p_j^2+
\omega^2 q_j^2+
 \big(q_{j+1}-q_j\big)^2 \right]\,,
\end{equation}
and periodic boundary conditions. 
For $\omega=0$ the chain is critical and its continuum limit is the $c=1$ free boson.
The construction of the partial transpose is detailed in \cite{hc1} 
and here we limit to presenting numerical checks of our CFT predictions. 
For $\omega=0$, the zero mode leads to divergent expressions, thus 
we work at finite but small $\omega$ such that $\omega L\ll 1$.

\begin{figure}[t]
\includegraphics[width=.48\textwidth]{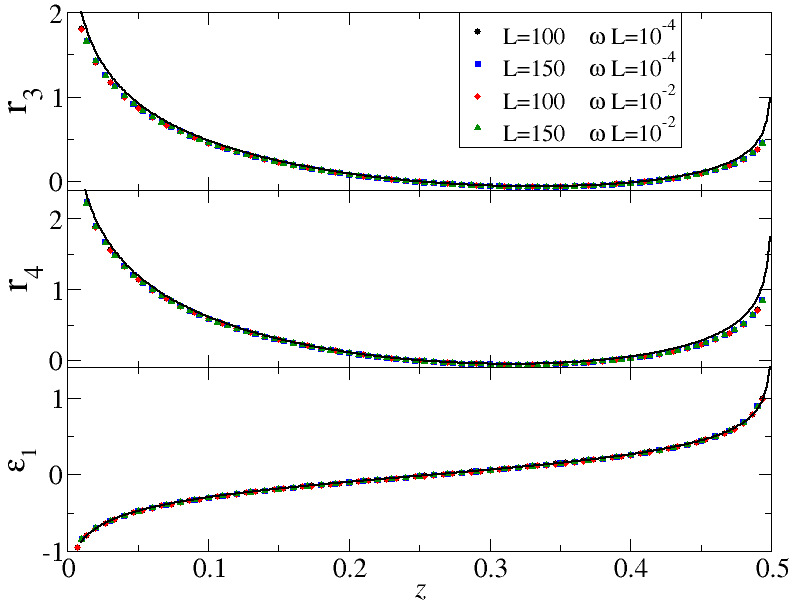}
\caption{
For two adjacent intervals of equal length $\ell<L/2$, we plot $r_n=\ln [\Tr(\rho_A^{T_{A_2=\ell}})^n/\Tr(\rho_A^{T_{A_2=L/4}})^n]$ 
as function of $z=\ell/L$. The subtraction is chosen  to cancel non-universal factors. 
The bottommost panel shows ${\cal E}_1={\cal E}-(\ln L)/4$ 
in which  non-universal terms are absent.
The continuous lines are the parameter free CFT predictions. 
}
\label{3ptfig}
\end{figure}

We first consider the case of two adjacent intervals of equal length $\ell$.
The results for $\Tr(\rho_A^{T_{2}})^n$ for $n=3,4$, as well as the results for the negativity ${\cal E}$
are reported in Fig.~\ref{3ptfig} where they are compared with the finite size CFT predictions
finding excellent agreement.

\begin{figure}[t]
\includegraphics[width=.48\textwidth]{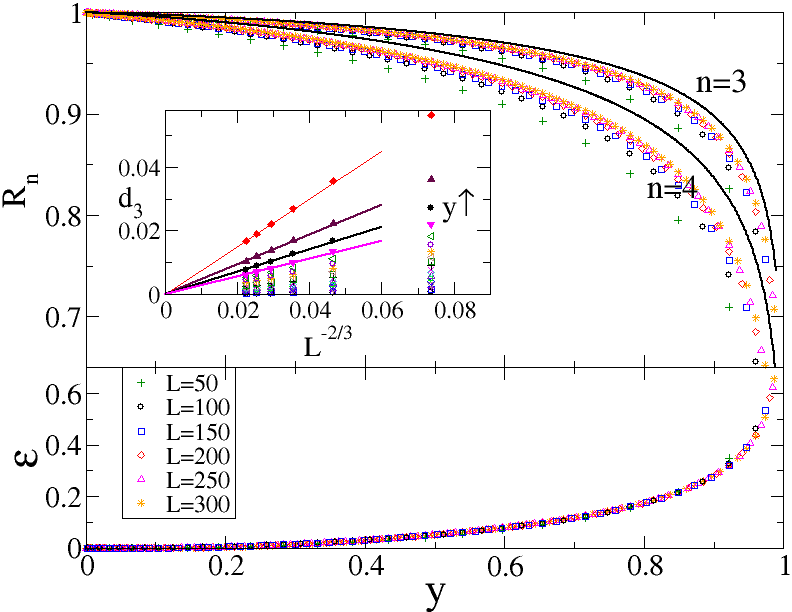}
\caption{Top: The ratio $R_n(y)$ in Eq. (\ref{Rndef}) as function of $y$ for several $L$ and for $n=3,4$.
The continuous lines are the parameter free CFT predictions.
The inset shows a finite size scaling analysis for $d_n\equiv R^{{\textrm{\tiny CFT}}}_n(y)-R_n(y)$ 
for $n=3$ displaying the unusual correction $L^{-2/n}$ \cite{unusual}. The same is true for higher $n$ \cite{cct-prep}.
Bottom: The negativity ${\cal E}(y)$ is a universal scale invariant function with an essential singularity at $y=0$. 
}
\label{4ptfig}
\end{figure}

The negativity of two disjoint intervals has been already considered numerically \cite{Neg2}.
We consider here the ratio
\be
R_n(y)\equiv \frac{\Tr(\rho_A^{T_{2}})^n}{\Tr\rho_A^n}\,,
\label{Rndef}
\ee
in which the non-universal parts due to the zero mode cancel and we are left with a universal function of $y$.
The CFT prediction for this ratio is \cite{cct-prep}
\be
R_n^{{\textrm{\tiny CFT}}}(y)=
\left[\frac{(1-y)^{\frac{2}{3}(n-\frac{1}{n})}  \prod_{k=1}^{n-1}  
F_{\frac{k}{n}}(y)  F_{\frac{k}{n}}(1-y)}{\prod_{k=1}^{n-1}  
\textrm{Re}\big(F_{\frac{k}{n}}(\tfrac{y}{y-1}) \bar{F}_{\frac{k}n}(\tfrac{1}{1-y}) \big)}\right]^{\frac12},
\label{RnCFT}
\ee
where $F_{q}(x)\equiv\! _2F_1(q,1-q,1,x)$ being $_2F_1(a,b,c,z)$ the hypergeometric function. 
This prediction is compared to the numerical data in Fig.~\ref{4ptfig}.
As $L$ increases, the data approach the CFT result. 
The differences with the asymptotic formula are due to the 
presence of 
unusual corrections to the scaling \cite{unusual} of the form $L^{-2/n}$.
A quantitative finite size scaling analysis  is reported in the inset of the figure. 
The bottom panel of Fig.~\ref{4ptfig} shows the negativity ${\cal E}$ for which all data collapse on a single curve, without 
sizable corrections. 
For small $y$, the data are very close to zero and are consistent with the form
$e^{-a/y}$ \cite{Neg2}, vanishing faster than any power.
For $y\to1$, we find $e^{{\cal E}(y)}\sim (1-y)^{-{1}/4} |\ln (1-y)|^{-1/2}$ as obtained
from the  analytic continuation of Eq.~(\ref{RnCFT}) in this regime \cite{cct-prep}. 
The logarithmic correction may be responsible for the 
exponent $\frac13$ found in Ref.~\cite{Neg2} 
as compared with our analytic result $\frac14$, which is consistent with our general result $\frac c4$.

{\it Conclusions}.
We described a general QFT formalism to calculate the logarithmic negativity. 
For a conformal invariant theory we worked it out for two intervals, both adjacent and disjoint. 
In the latter case, the negativity is a universal scale invariant function. 
Some generalizations  such as for compactified free boson, Ising CFT, finite temperature CFT, and massive QFT 
have been already obtained  and will be presented elsewhere \cite{cct-prep}.

However, there are still open problems, among them the analytic continuation  $n_e\to1$ of the results for 
disjoint intervals which remains a formidable  task,
reflecting a similar problem for the  entanglement entropy \cite{cct-09,cct-11}.

Finally, it is of extreme interest to check numerically our CFT predictions  in more complicated 
lattice models such as spin-chains and itinerant fermions. 

{\it Acknowledgments}. ET thanks Marcus Cramer for discussions. 
This work was supported by the 
ERC under  Starting Grant  279391 EDEQS (PC). 
This work has been partly done when the authors were guests of the Galileo Galilei Institute in Florence 
and Institut Henri Poincar\'e in Paris.


\begin{thebibliography}{99}


\bibitem{rev}
L. Amico, R. Fazio, A. Osterloh, and V. Vedral, 
Rev. Mod. Phys, {\bf 80}, 517 (2008); 
J. Eisert, M. Cramer, and M. B. Plenio, 
ibid. {\bf 82}, 277 (2010);
Entanglement entropy in extended systems, 
P.~Calabrese, J.~Cardy, and B. Doyon, J. Phys. A {\bf 42}, 500301 (2009).


\bibitem{vw-01}
G. Vidal and R. F. Werner, Phys. Rev. A {\bf 65}, 032314 (2002).


\bibitem{fazio}
A. Osterloh, L. Amico, G. Falci, and R. Fazio, 
Nature {\bf 416}, 608 (2002).


\bibitem{Holzhey} C. Holzhey, F. Larsen, and F. Wilczek,
Nucl. Phys. B {\bf 424}, 443 (1994).

\bibitem{Vidal}
G. Vidal, J. I. Latorre, E. Rico, and A. Kitaev,
Phys. Rev. Lett. {\bf 90}, 227902 (2003);
J. I. Latorre, E. Rico, and G. Vidal,
Quantum Inf. Comput. {\bf 4}, 048 (2004).

\bibitem{cc-04} P.~Calabrese and J.~Cardy, J. Stat. Mech. P06002 (2004);
J. Phys. A  {\bf 42}, 504005 (2009).

\bibitem{cct-09}
P.~Calabrese, J. Cardy, and E. Tonni, J. Stat. Mech P11001 (2009).

\bibitem{cct-11}
P.~Calabrese, J. Cardy, and E. Tonni, J. Stat. Mech P01021 (2011).

\bibitem{fps-08}
S. Furukawa, V. Pasquier, and J. Shiraishi, Phys. Rev. Lett. {\bf 102}, 170602 (2009);
M. Caraglio and F. Gliozzi, JHEP 0811: 076 (2008);
H. Casini and M. Huerta, JHEP 0903: 048 (2009);
V. Alba, L. Tagliacozzo, and P. Calabrese, Phys. Rev. B, {\bf 81} 060411  (2010); J. Stat. Mech. (2011) P06012;
M. Fagotti and P. Calabrese,  J. Stat. Mech. (2010) P04016;
M. Fagotti,  EPL {\bf 97}, 17007 (2012).

\bibitem{Neg1}
H. Wichterich, J. Molina-Vilaplana, and S. Bose, Phys. Rev. A {\bf 80}, 010304 (2009). 

\bibitem{Neg2}
S. Marcovitch, A. Retzker, M. B. Plenio, and B. Reznik, Phys. Rev. A {\bf 80}, 012325 (2009).

\bibitem{Neg3}
H. Wichterich, J. Vidal, and S. Bose,  Phys. Rev. A {\bf 81}, 032311 (2010).

\bibitem{lev}
K. Zyczkowski, P. Horodecki, A. Sanpera, and M. Lewenstein,
Phys. Rev. A {\bf 58}, 883 (1998).

\bibitem{klein}
M. Bianchi, G. Pradisi, and A. Sagnotti, Nucl.Phys. B {\bf 376}, 365 (1992).

\bibitem{cft-book}
P. Di Francesco, P. Mathieu, and D. Senechal, Conformal Field Theory (Springer-Verlag, New York, 1997).

\bibitem{cct-prep}
P. Calabrese, J. Cardy, and E. Tonni, in preparation.

\bibitem{hc1}
K. Audenaert, J. Eisert, M. B. Plenio and R. F. Werner,  Phys. Rev. A {\bf 66}, 042327 (2002).


\bibitem{unusual}
P. Calabrese, M. Campostrini, F. Essler, and B. Nienhuis, Phys. Rev. Lett. {\bf 104}, 095701 (2010); 
J. Cardy and P. Calabrese, J. Stat. Mech. (2010) P04023;
P. Calabrese and F. H. L. Essler, J. Stat. Mech. (2010) P08029.  









\end{thebibliography}
\end{document}